\documentclass[utf8]{frontiersFPHY} 

\usepackage{url,hyperref,lineno,microtype,subcaption}
\usepackage[onehalfspacing]{setspace}

\def\firstAuthorLast{Jose P. Rodriguez {et~al.}} 
\def\Authors{Jose P. Rodriguez\,$^{1,*}$, Dmytro S. Inosov\,$^{2}$ and Jun Zhao\,$^{3}$}
\def\Address{$^{1}$ Department of Physics and Astronomy, California State University at Los Angeles,
 Los Angeles, California, USA \\
$^{2}$ Institute of Solid State and Materials Physics, Technische Universit\"at Dresden, 01069 Dresden, Germany  \\
$^{3}$ Department of Physics, Fudan University, Shanghai, China}
\def\corrAuthor{Jose P. Rodriguez, Department of Physics and Astronomy, California State University at Los Angeles,
Los Angeles, California 90032, USA}

\def\corrEmail{jrodrig@calstatela.edu}

\begin{document}
\onecolumn
\firstpage{1}

\title[Editorial  ...]{Editorial on Research Topic:
High-Tc Superconductivity in Electron-Doped Iron Selenide and Related Compounds}

\author[\firstAuthorLast ]{\Authors} 
\address{} 
\correspondance{} 

\extraAuth{}

\maketitle




Iron-selenide superconductors comprise a particularly interesting group of materials
 inside the family of iron-based superconductors.
The simplest member of the group is bulk FeSe, which has a modest critical temperature of $T_c = 9$ K.
Like iron-pnictide superconductors, bulk FeSe shows a structural transition at $T_s = 90$ K from a 
tetragonal to an orthorhombic phase driven by nematic ordering of the electronic degrees of freedom.
Angle-resolved photoemission spectroscopy (ARPES), for example,
 reveals a small hole Fermi surface pocket at the
center of the Brillouin zone and two electron Fermi surface pockets at the corner of the Brillouin zone,
each with unequal $d_{xz}/d_{yz}$-orbital character.
Unlike in iron-pnictide superconductors, however, no magnetic order coexists with the nematic order
at temperatures below the structural transition.
Inelastic neutron scattering (INS) spectroscopy
 finds a spin resonance inside the energy gap
of the superconducting phase in bulk FeSe, however, at wavevectors
corresponding to a stripe  spin-density wave (SDW)\cite{wang_zhao_16a}.
It  strongly suggests $s^{+-}$ superconductivity across the hole and electron Fermi surface pockets
driven by associated antiferromagnetic spin fluctuations.
INS also finds spin fluctuations at  the N\'eel wavevector $(\pi,\pi)$
above the superconducting energy gap\cite{wang_zhao_16b}.
This suggests that
superconductivity, nematic order, stripe-SDW order,
and some type of N\'eel antiferromagnetic order compete at low temperature in bulk FeSe. 
One of the editors of the research topic has proposed that the latter is  hidden N\'eel order\cite{jpr_20,jpr_21}.

The superconducting critical temperature increases dramatically to
$30$-$40$ K and above upon doping iron selenide with electrons.
The latter has been achieved in various ways;
 for example, by alkali-metal intercalation, by placing a monolayer of FeSe
on a substrate, and by organic-molecule intercalation.
ARPES finds that the hole bands at the center of the Brillouin zone lie buried below the Fermi level.
INS finds a spin resonance inside the superconducting energy gap,
but it lies midway between the SDW and N\'eel wavenumbers\cite{park_keimer_inosov_11}.
INS also finds peaks and rings of low-energy spin excitations above the energy gap 
around the N\'eel wavevector\cite{friemel_keimer_inosov_12,pan_zhao_17}.
ARPES and scanning tunneling microscopy (STM) find a non-zero superconducting energy gap.
The situation with electron-doped FeSe  is rather puzzling then,
with high-$T_c$ superconductivity existing  over electron Fermi surface pockets alone!
This is not expected in iron-selenide superconductors, 
where electron-electron repulsion is strong\cite{yi_15}.
The latter requires that the sign of the pair wavefunction oscillates over the Brillouin zone\cite{jpr_21}.

It is our pleasure to introduce eight articles from the Research Topic that address many of
the unsolved problems that have emerged in the field of iron-selenide superconductors,
some of which we have listed above.
The contributions to the Research Topic contain articles on  both theory and experiment,
with four papers reporting on original research,
and with four review papers.
Tong Chen, Min Yi and Pengcheng Dai review how nematicity in bulk iron selenide
can be scrutinized by exploiting detwinning techniques\cite{rice_u}, 
while Amalia Coldea reviews the series of nematic superconductors FeSe$_{1-x}$S$_x$\cite{coldea}.
Both articles tackle the interplay between nematicity and superconductivity that exists in bulk FeSe,
with or without chemical substitutions.
Maw-Kuen Wu and collaborators show that insulating  Fe$_4$Se$_5$ becomes a superconductor with $T_c = 8$ K 
after proper annealing\cite{wu_20}.  
They thereby argue that Fe$_4$Se$_5$ is the insulating  parent compound for  iron-selenide superconductors.
It would clearly be useful to compare future studies of the low-energy spin excitations
in Fe$_4$Se$_5$ with those of its electron-doped counterpart
 Rb$_2$Fe$_4$Se$_5$ \cite{park_keimer_inosov_11,friemel_keimer_inosov_12}.
Finally, Xiaoli Dong, Fang Zhou and Zhongxian Zhao review a new soft-chemical technique to
grow high-quality  single crystals of organic-molecule intercalated FeSe\cite{cas}. 
Their samples have critical temperatures of $T_c = 42$ K,
and they notably show record critical currents.

On the theory side, Rong Yu, Qimiao Si and collaborators 
review $3d$-orbital-selective physics in iron superconductors\cite{yu_si}.
They point out how the $d_{xy}$  orbital is 
the one most susceptible to Mott localization in iron-selenide superconductors\cite{yi_15}.
They also emphasize how
the relatively small energy splitting between the $d_{xz}/d_{yz}$ orbitals 
that is seen by ARPES in the nematic phase of bulk FeSe,
$\Delta E_{\Gamma}$ and $\Delta E_M < 50$ meV,
 can be reconciled with the large orbitally-dependent
wavefunction renormalizations seen by STM in the same phase,
$Z(d_{yz}) / Z(d_{xz}) = 4$.
Maxim Dzero and Maxim Khodas study the effect of point disorder on the stripe SDW state
by exploiting a quasi-classical Green's function technique\cite{dzero_khodas}.
They find that the tetragonally symmetric stripe SDW state is more robust with respect to disorder
than the orthorhombically symmetric one.
This result could have bearing on the absence of magnetic order in the nematic phase of bulk FeSe, for example.
Last, Andrzej Ptok, Konrad Kapcia and Przemysław Piekarz 
study a two-band model for iron superconductors 
that includes intra-band and inter-band coupling between Cooper pairs\cite{poles}.
They notably find Cooper pair states in relative orbitals of mixed symmetry.
Finally,
Rustem Khasanov and collaborators applied muon-spin rotation/relaxation ($\mu$SR)
on the iron-pnictide superconductor NdFeAsO$_{0.65}$F$_{0.35}$,
thereby obtaining London penetration lengths\cite{gupta_khasanov}.
Interestingly,
a two-band analysis of their data yields only weak inter-band coupling of the Cooper pairs.

The brief survey above of the author contributions to the Research Topic conveys
the richness of the field of iron-selenide superconductivity and related materials.
We believe that you will enjoy reading the Research Topic.

Yours sincerely, Jose Rodriguez, Dmytro Inosov and Jun Zhao,
February 28, 2022.

\end{document}